\documentclass{ws-ijmpa}
\usepackage[super,compress]{cite}
\usepackage{graphicx}
\usepackage{hyperref}
\hypersetup{colorlinks,urlcolor=black,citecolor=black,linkcolor=black,filecolor=black}
\usepackage{breakurl}

\newcommand{\SU}{\mathrm{SU}}
\newcommand{\U}{\mathrm{U}}

\newcommand{\Tr}{{\rm Tr}}
\newcommand{\tr}{{\rm Tr}}
\newcommand{\real}{{\rm Re\,}}

\newcommand{\cut}[1]{}
\newcommand{\AB}{\theta_{\mbox{\tiny{H}}}}
\newcommand{\Veff}{V_{\mbox{\tiny{eff}}}}
\newcommand{\LambdaR}{\Lambda_{\mbox{\tiny{R}}}}
\newcommand{\LambdaUV}{\Lambda_{\mbox{\tiny{UV}}}}
\newcommand{\comma}{\; ,}

\begin{document}
\markboth{F.~Knechtli and E.~Rinaldi}{Extra-dimensional models on the lattice}

%
\catchline{}{}{}{}{}
%

\title{Extra-dimensional models on the lattice}

\author{Francesco Knechtli}

\address{Department of Physics, Bergische Universit\"at Wuppertal\\
Gaussstr. 20, D-42119, Wuppertal, Germany\\
knechtli@physik.uni-wuppertal.de}

\author{Enrico Rinaldi}

\address{Nuclear and Chemical Sciences Division\\ Lawrence Livermore National Laboratory\\
Livermore, CA 94550, USA\\
rinaldi2@llnl.gov}

\maketitle

\begin{history}
\received{13 May 2016}
\end{history}

\begin{abstract}
In this review we summarize the ongoing effort to study extra-dimensional
gauge theories with lattice simulations.
In these models the Higgs field is identified with extra-dimensional components
of the gauge field. The Higgs potential is generated by quantum corrections
and is protected from divergencies by the higher dimensional gauge symmetry.
Dimensional reduction to four dimensions can occur through compactification or
localization. Gauge-Higgs unification models are often studied using
perturbation theory.
Numerical lattice simulations are used to go beyond these perturbative
expectations and to include non-perturbative effects.
We describe the known perturbative predictions and their fate in the
strongly-coupled regime for various extra-dimensional models.
\begin{flushright}
WUB/16-01, LLNL-JRNL-691819
\end{flushright}

\end{abstract}


\section{Gauge-Higgs unification: a scalar from extra dimensions}
\label{sec:intro}

The Higgs sector of the Standard Model is the only one containing an
elementary scalar field.
After the discovery of the Higgs boson~\cite{ATLAS:2012gk,CMS:2012gu}
and the measurements of its
couplings at the LHC in recent years, we know that the Higgs sector is
a remarkably good effective description of low-energy electro-weak
physics.
However, we would like to explain why the Higgs mass is so light and
we would like to know if the spontaneous symmetry breaking (SSB) mechanism
described by Brout, Englert and Higgs~\cite{Englert:1964et,Higgs:1964ia,Higgs:1964pj}
has a more profound origin,
maybe from a more complete theory at high energy.
The recent observation of a diphoton excess in the scalar channel at a
bout $750$ GeV at the ATLAS and CMS
experiments~\cite{CMS:2015dxe,ATLAS:2015hp}, if confirmed, will find
no explanation within the Standard Model, bringing forth the need for
a more general theory.

Intriguingly, the existence of hidden extra
dimensions~\cite{Kaluza:1921tu,Klein:1926tv} can address
both the lightness of the Higgs and the origin of SSB, while providing
a natural framework for the unification of gauge forces.
The idea that the Higgs field may be directly related to the
components of extra-dimensional gauge fields was first discussed in
Ref.~\citen{Manton:1979kb} and it is the foundational concept thorough
all the models we discuss in this review.

Five-dimensional Yang--Mills theories compactified on a circle have a
light scalar mode, whose mass renormalization $m_5$ in perturbation theory
is protected by the remnant of the higher-dimensional gauge
symmetry~\cite{Hosotani:1983xw,Hatanaka:1998yp,Cheng:2002iz,vonGersdorff:2002as,Hosotani:2005fk,Hosotani:2007kn,DelDebbio:2008hb}.
The extra-dimensional gauge theory is naively non-renormalizable and
is regulated with a cutoff $\LambdaUV$, which interestingly does not
affect the quantum corrections to the scalar mass at one loop:
\begin{equation}
  \label{eq:one-loop-mass}
  m_5^2 \; = \; \frac{9 g_5^2 N_c}{32 \pi^5 R^3} \zeta (3)
  \comma
\end{equation}
where $g_5$ is the five-dimensional gauge coupling and $\zeta$ is the
Riemann Zeta-function.~\footnote{
Eq.~\eqref{eq:one-loop-mass} applies to a circular extra dimension. If
it is an interval (orbifold), there is an extra factor $1/2$.}

\begin{figure}[t]
\centerline{\includegraphics[scale=0.25]{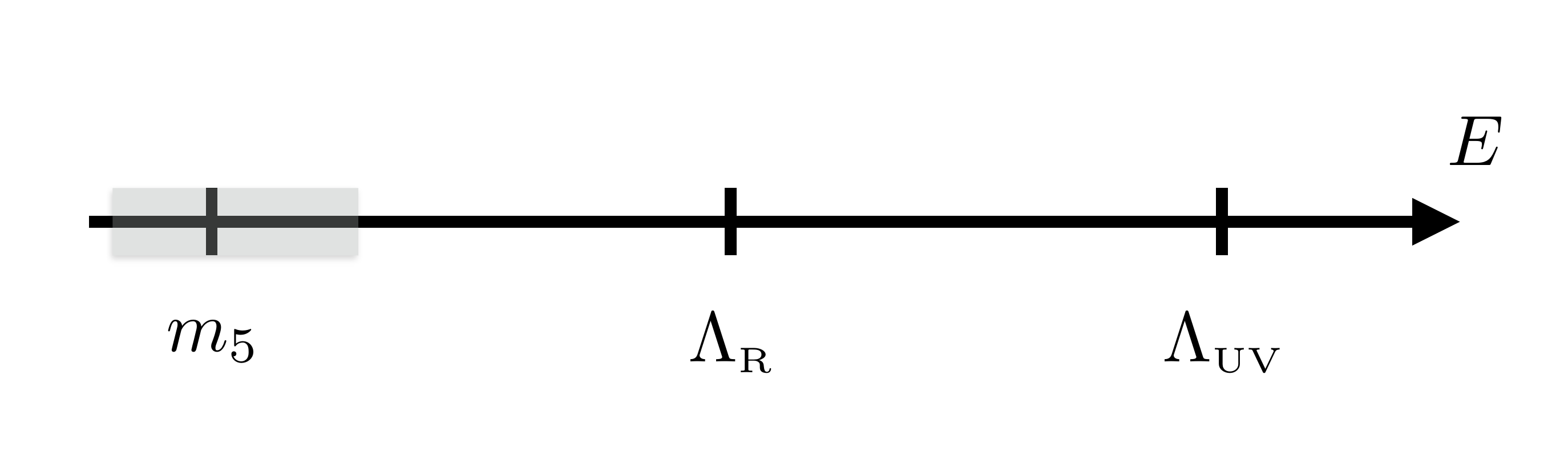}}
\caption{The figure shows the desired separation of energy scales in a
  five-dimensional theory with cutoff $\LambdaUV$ and compactification
  scale $\LambdaR=\frac{1}{2\pi R}$.
  The shaded area is the region where we expect the scalar mass $m_5$
  to be described by Eq.~\eqref{eq:one-loop-mass}. \label{fig:scale-sep}}
\end{figure}

This relation is true if the details of the regularization can be
neglected at the scale of compactification,
$\LambdaR \sim R^{-1} \ll \LambdaUV$, but its validity is restricted to the
weak-coupling region.
Lattice simulations are able to access non-perturbative physics,
and numerically investigate the theory in the regime where
the hierarchy of scales is such that the low-energy physics is
described by a four-dimensional effective theory with a light
scalar particle and check if Eq.~\eqref{eq:one-loop-mass} holds.
The desired separation of scales is pictorially shown
in Fig.~\ref{fig:scale-sep}.
This scenario is summarized in Sec.~\ref{sec:pure-gauge-theory}.

The scalar particle that is present in the dimensionally reduced
theory is in the same representation of the gauge group as the
extra-dimensional gauge field.
Therefore, the compactification mechanism broadly described above for
a pure gauge theory will provide an adjoint scalar, which can
not describe the Higgs particle.
A fundamental scalar can be obtained if the extra
dimension is compactified on an orbifold instead of a circle.
This scenario is reviewed
in Sec.~\ref{sec:non-pert-gauge} and has the additional advantage to
provide a description for chiral fermions from an extra dimension.

In perturbation theory, Eq.~\eqref{eq:one-loop-mass} arises from an
effective potential that is generated by radiative correction.
This effective potential has a trivial minimum in a Yang--Mills
theory, but is modified by the addition of fermions in the adjoint or
in the fundamental representation.
The fermionic effects destabilize the trivial minimum and a vacuum
that breaks the gauge symmetry is possible.
This is the basis of the Hosotani mechanism~\cite{Hosotani:1983xw}
whose non-perturbative application is explored in
Sec.~\ref{sec:ferm-hosot-mech}.
There exists an alternative non-perturbative mechanism of spontaneous
gauge symmetry breaking. It takes place on the orbifold (but not on the torus)
and relies on the SSB of a global symmetry in accordance with
Elitzur's theorem\cite{Elitzur:1975im}. This will be discussed in
Sec.~\ref{sec:non-pert-gauge}.

Dimensional reduction to four dimensions can occurr if the extra dimension
becomes compact. This is the Kaluza--Klein mechanism. Another mechanism is
localization. In flat space localization can be realized in a way similar to
the layered phase~\cite{Fu:1983ei}.
At present, the searches that the LHC performs for discovering extra 
dimensions are connected to the compactification scenario.
It is possible  that if there exists an extra dimension with an
underlying localization  mechanism at work, the experimental
signatures may be different.

The review is structured as follows.
In Sec.~\ref{sec:lattice} we introduce the general setting used to
study extra-dimensional models on the lattice
by numerical Monte Carlo simulations and in
mean-field analytical calculations that we also introduce in the same section.
Sec.~\ref{sec:pure-gauge-theory} is devoted to a summary of results
for Yang--Mills theories in more than four dimensions with toroidal
boundary conditions.
Then we discuss theories with fermions and the accompanying Hosotani
mechanism in Sec.~\ref{sec:ferm-hosot-mech}.
Finally, we report the results of the study of Yang--Mills
theories with orbifold boundary conditions and the non-perturbative
realization of gauge-Higgs unification in Sec.~\ref{sec:non-pert-gauge}.

We do not discuss extra-dimensional supersymmetric models or
the importance of extra dimensions in string theories and other
theories of gravity.
Lattice studies in those directions are covered by other reviews of
this series (Supersymmetry and Gauge/Gravity duality).

\section{Extra-dimensional gauge theories on the lattice}
\label{sec:lattice}

Lattice methods are useful to investigate the phase diagram of
extra-dimensional gauge theories and to study the non-perturbative
viability of various mechanisms for dimensional reduction. In order to
have a self-contained description, in this section we summarize different
geometries studied on the lattice and, in later
sections, we will describe their emerging properties and phase diagrams.

\subsection{Torus geometry}
\label{sec:torus-geometry}

The original mathematical idea of extra
dimensions~\cite{Kaluza:1921tu,Klein:1926tv} relies on the existence of
a compactification radius $R$. On a torus geometry, where periodic
boundary conditions  for the gauge fields are imposed in all
directions, compactification can be easily achieved by reducing the size of
the extra dimension $L_5=2\pi R$ below a critical value $L_{5c}$ and
keeping the other dimensions large.

It is rather straightforward to study this compactified theory on the
lattice.
As a first step we discretize the
continuum five-dimensional Euclidean action using the Wilson plaquette
action. For an $\SU(N_c)$ gauge theory on the torus this is given
by~\cite{Ejiri:2000fc}
\begin{eqnarray}\label{e:torusaction}
S_W^{tor} & = & \sum_{n_{\mu}} \sum_{n_{5}=0}^{N_{5}-1} {\left[ 
\frac{\beta_{4}}{N_c}  \sum_{\mu<\nu}
{\real\tr\left\{1 - U_{\mu\nu}(n_{\mu},n_{5}) \right\}} + \right.} \nonumber \\
& & {\left. \frac{\beta_{5}}{N_c} \sum_\mu
{\real\tr\left\{ 1 - U_{\mu5}(n_{\mu},n_{5}) \right\}} \right]}~,
\end{eqnarray}
where $\beta_{4}$ and $\beta_{5}$ denote the lattice coupling in the four 
standard dimensions and in the fifth (extra) dimension
respectively, $U_{MN}(n_{\mu},n_{5})$ with $M,N=0,1,2,3,5$ is the plaquette in 
directions $M$ and $N$ located on lattice site
indexed by $n_{\mu}$ in the four standard dimensions and $n_{5}$ in the extra 
dimension. This is a standard choice~\cite{Ejiri:2002ww,Irges:2009bi,Farakos:2010ie,deForcrand:2010be,Knechtli:2011gq,DelDebbio:2012mr}.

The action Eq.~(\ref{e:torusaction}) with the lattice couplings $\beta_4$ and 
$\beta_5$ allows for different lattice spacings 
$a_{4}$ and $a_{5}$, where $a_{4}$ 
denotes the lattice spacing in the usual four dimensions and $a_{5}$ denotes 
the lattice spacing in the extra dimension. This is naturally
suited for choosing the size of the extra dimension $L_5=2\pi R=N_5 a_5$
independently from the size of the four dimensional space-time $L =
N_4 a_4$. Here $N_4$ and $N_5$ are the number of lattice points in each of the
four directions and in the fifth direction, respectively.

An equivalent parametrization for Eq.~(\ref{e:torusaction}) uses
$\beta=\sqrt{\beta_4\beta_5}$ and
$\gamma=\sqrt{\beta_5/\beta_4}$.
In the classical limit $(a_4,\ a_5) \to0$
\begin{equation}\label{e:acouplings}
\gamma = \frac{a_{4}}{a_{5}} \quad\mbox{and}\quad 
\beta = \frac{2N_c a_{4}}{g^{2}_{5}}\,, 
\end{equation}
where $g_{5}$ is the dimensionful continuum gauge coupling. Therefore,
$\gamma$ is often called the \emph{anisotropy} coefficient, giving a
tree-level prediction for the ratio of lattice spacings. The
non-perturbative anisotropy $\xi$ can be directly measured by looking at
ratios of lattice correlation functions~\cite{Ejiri:1998xd}.

The regime where $\gamma > 1$ is clearly associated with a small spacing in the
extra dimension $a_5$ compared to $a_4$.
The size of the extra dimension with respect to the
four-dimensional lattice spacing can be written as $\tilde{N}_5=N_5/\gamma$.
On the other hand, the regime where $\gamma < 1$ is associated with a larger lattice
spacing along the extra dimension than in the four-dimensional hyperplanes
orthogonal to it.
In the limit $\beta_5=0$ the action
Eq.~\eqref{e:torusaction} describes $N_5$ independent copies of 
four-dimensional gauge theories on the hyperplanes. 
The regime at $\gamma < 1$ is connected with this limit.

As a second step in the study of extra-dimensional theories on the
lattice, different values for dimensions $(L,L_5)$ and the gauge coupling
$g_5^2$ of the continuum theory can be probed by
modifying the parameters of the lattice model
$(N_4,N_5,\beta_4,\beta_5)$.
In the lattice model's parameter space, physical properties of the low
energy regime of the theory, such as a
mass gap or the static potential, are studied with traditional
lattice methods described in the Introduction by the editors.

We should note that five-dimensional gauge theories are perturbatively
non-renormalizable, 
due to the negative mass dimension of the gauge coupling $g_5$.
In the continuum
if the cutoff of the theory is infinite the coupling is necessarily zero.
On the lattice the cutoff is given by the inverse lattice spacing $a_4^{-1}$.
The only continuum limit $a_4 \to0$, 
which is known to exist so far, is the perturbative trivial limit.
Triviality can also be understood by considering the one-loop renormalization of
the effective four-dimensional coupling
$g_4^2=g_5^2/(N_5a_5)$~\cite{Dienes:1998vg,Irges:2007qq}.

\subsection{Orbifold geometry}
\label{sec:orbifold-geometry}

The orbifold theory we consider here is defined in the five-dimensional
domain $I = \{ n_{\mu}, 0 \leq n_{5} \leq N_{5} \}$ with 
volume $N_t \times N_s^{3} \times N_{5}$.
The Wilson action for an $\SU(N_c)$ gauge theory on this
orbifold is given by~\cite{Irges:2004gy,Knechtli:2005dw}
\begin{eqnarray}\label{e:orbiaction}
S_W^{orb} & = & \sum_{n_{\mu}} {\left[ 
\frac{\beta_{4}}{N_c} \sum_{n_{5}=0}^{N_{5}} \sum_{\mu<\nu}
{w~\real\tr\left\{1 - U_{\mu\nu}(n_{\mu},n_{5}) \right\}} + \right.} \nonumber\\
&& {\left. \frac{\beta_{5}}{N_c}  \sum_{n_{5}=0}^{N_{5}-1} \sum_\mu
{\real\tr{\left\{ 1 - U_{\mu5}(n_{\mu},n_{5}) \right\}}} \right]}~,
\end{eqnarray}
which follows the parametrization of Eq.~(\ref{e:torusaction}).
The weight $w$ is due to the orbifold geometry and is set to 
$\frac{1}{2}$ for plaquettes on the boundary and $1$ elsewhere.
Lattice gauge fields on the orbifold obey the boundary conditions
\begin{equation}\label{e:orbibc}
U_\mu(n_\mu,n_5) = g~U_\mu(n_\mu,n_5)~g^{-1} \,,\quad\mbox{for $n_5=0,N_5$} \,.
\end{equation}
The matrix
$g$ is an element of $\SU(N_c)$ such that $g^2$ is in the center $\mathbb{Z}_{N_c}$
of $\SU(N_c)$.
The boundary conditions Eq.~(\ref{e:orbibc}) break in general the gauge group
$\SU(N_c)$ down to a subgroup $H$ on the orbifold boundaries according to the 
pattern $SU(p+q)\longrightarrow SU(p)\times SU(q)\times U(1)$ (see for example 
Ref.~\citen{Hebecker:2001jb}).
As an example, for gauge group $\SU(2)$ the choice $g=-i\sigma^3$ breaks the
gauge group to $\U(1)$ on the orbifold boundaries.

The operators representing the scalar and vector particles on the
orbifold are defined as follows. 
The left-to-right boundary-to-boundary-line is denoted by 
$l(n_\mu) = \prod_{n_5=0}^{n_5=N_5-1}U_5(n_\mu,n_5)$.
Starting from the Polyakov loop on the torus, its orbifold projection yields
$p=l\, g\, l^\dagger\, g^\dagger$, which is
a field living on the left boundary (an analogous field living
on the right boundary can also be defined).
Scalar operators can be defined as 
\begin{equation}\label{e:Higgsorb}
{\cal P}=\mathrm{tr}~(p) \quad\mbox{or}\quad 
{\cal H}=\mathrm{tr}~(hh^\dagger)
\end{equation}
using for $h$ the expression
\begin{equation}\label{e:gaugepotorb}
h(n_\mu) = \frac{1}{4N_5} [p(n_\mu) - p^\dagger(n_\mu), g]~.
\end{equation}
Vector gauge boson operators can be defined as
\begin{equation}\label{e:gaugebosonorb}
{\cal Z}(n_\mu) = \mathrm{tr}~\left(
g~U_k(n_\mu,n_5=0)~h(n_\mu+\hat{k})~U_k(n_\mu,n_5=0)^\dagger~h(n_\mu)\right)\,.
\end{equation}
These operators were introduced in Ref.~\citen{Irges:2006hg}.

The theory defined by Eq.~(\ref{e:orbiaction}) possesses a global
stick symmetry, $\mathcal{S}$~\cite{Ishiyama:2009bk}. This symmetry
is defined by the combination
$\mathcal{S} = \mathcal{S}_{L} \cdot \mathcal{S}_{R}$, where
$\mathcal{S}_{L}$ is a symmetry defined on the left boundary via
\begin{equation}\label{e:stickleft}
U_5(n_{5} = 0) \rightarrow g_{s}^{-1}~U_5(n_{5} = 0)
\quad\mbox{and}\quad
U_\nu(n_{5} = 0) \rightarrow g_{s}^{-1}~U_\nu(n_{5} = 0)~g_{s}~. 
\end{equation}
(We suppress here the coordinate $n_\mu$.)
The symmetry $\mathcal{S}_{L}$ is generated by an element $g_{s}$ with
$\{g,g_{s}\} = 0$,
which is not a gauge or center transformation.
$g_{s}$ is an element of the generalized
Weyl group, $W_{\SU(N_c)}(H) = N_{\SU(N_c)}(H)/H$, which is the coset of the 
normalizer of $H$ in $\SU(N_c)$ divided by $H$~\cite{Irges:2013rya}.
As an example, for the $\SU(2)$ orbifold with $g=-i\sigma^3$, a stick
matrix is $g_s=-i\sigma^1$.
$\mathcal{S}_{R}$ can be defined on the right boundary in an equivalent fashion.
\begin{equation}\label{e:stickright}
U_5(n_{5} = N_5-1)   \rightarrow U_5(n_{5} = N_5-1)~g_{s}
\quad\mbox{and}\quad
U_\nu(n_{5} = N_5) \rightarrow g_{s}^{-1}~U_\nu(n_{5} = N_5)~g_{s}~. 
\end{equation}
The scalar operators in Eq.~(\ref{e:Higgsorb}) are invariant under
the stick symmetry $\mathcal{S}$ while the gauge boson operator in
Eq.(\ref{e:gaugebosonorb}) is odd under this symmetry. This latter property
plays an important role as we will see in Sec.~\ref{sec:npghu-mc-results}.

\subsection{Mean-field approach}
\label{sec:mean-field-approach}

The mean-field approach to lattice gauge theory is 
reviewed in Ref.~\citen{Drouffe:1983fv}. We briefly introduce it here.
The link variables $U$ in $\SU(N_c)$
are traded for the unconstrained complex variables $V$ and the
Lagrange multipliers $H$ used to represent the $\delta$ functions 
$\delta(V-U)$. The partition function for gauge action $S_W$
can be rewritten as
\begin{eqnarray}
Z & = & \int {\cal D}[V] \int {\cal D}[H] \, e^{-S_{\rm eff}[V,H]} \,,\nonumber \\
S_{\rm eff}& = &S_W[V] + 
\sum_{n,M} \left[u(H_M(n)) + 
(1/N_c)\mathrm{Re}\,\mathrm{tr}\,\{H_M(n)V_M(n)\} \right] \label{e:effaction}\,
\end{eqnarray}
where the effective mean-field action $u(H)$ for a given link $U_M(n)$ 
is defined via
\begin{equation}\label{e:ueff}
\mathrm{e}^{-u(H)} = \int \mathrm{d}U_M(n) \, 
\mathrm{e}^{(1/N_c)\mathrm{Re}\,\mathrm{tr}\,\{U_M(n)H\}} \,.
\end{equation}
The mean-field or zeroth order approximation amounts to finding the minimum
of the effective action. On a periodic isotropic lattice the variables 
$V$ and $H$ are set proportional to the unit matrix, 
$H=\bar{H}~\mathbf{1}$ and $V=\bar{V}~\mathbf{1}$.
The zeroth order saddle point solution or "mean-field background" 
$\bar{H}$ and $\bar{V}$ can be easily obtained by 
taking derivatives of $S_{\rm eff}$ in Eq.~(\ref{e:effaction}) with respect to $V$ and $H$
and require them to vanish.
Gaussian fluctuations are defined by setting
$H = \bar{H} + h$ and $V = \bar{V} + v$. For the calculation of
corrections stemming from fluctuations gauge fixing is necessary.
Covariant gauge fixing on $v$ is a practical choice. 
In Ref.~\citen{Ruhl:1982er} 
it was shown that this is equivalent to gauge-fix the original links $U$.
We denote the quadratic part of the effective action by 
\begin{equation}\label{e:effaction2}
S^{(2)}[v,h]=\frac{1}{2}\left[
h^T K^{(hh)} h + 2v^T K^{(vh)} h + v^T (K^{(vv)} + K^{({\rm gf})}) v\right] \,,
\end{equation}
where $K^{({\rm gf})}$ is the contribution from the gauge fixing term.
The expectation value $\langle {\cal O}(U) \rangle$ of an observable
${\cal O}$ to first order in the mean-field expansion is given by
\begin{equation}\label{e:mfcorrection}
\langle {\cal O} \rangle  =
{\cal O}[{\overline V}] + \frac{1}{2}{\rm tr} 
\left\{\frac{\delta^2{\cal O}}{\delta V^2}\Biggr|_{\bar{V}} K^{-1}\right\} \,,
\end{equation}
where the propagator $K$ is defined as
\begin{equation}\label{e:mfinvprop}
K=-\left(K^{(hh)}\right)^{-1}+K^{(vv)}+K^{({\rm gf})}
\end{equation}
and the second derivative of the observable is evaluated in the mean-field
background.
Masses are calculated by evaluating
the connected two point point function $C(t)$ of a time dependent observable
${\cal O}(t)$ in the mean-field expansion.
To first order in the fluctuations the expression reduces to 
$C(t)=C^{(1)}(t)$ with
\begin{equation}\label{e:mfmass1}
 C^{(1)} (t) = \frac{1}{2} \mathrm{tr}\, 
\left\{ \frac{\delta^{(1,1)}{\cal O}(t_0+t){\cal O}(t_0) }{\delta^2 V } K^{-1}
\right\} \, ,
\end{equation}
where the notation $\delta^{(1,1)}$ means one derivative acting on each 
of the ${\cal O}(t_0+t)$ and ${\cal O}(t_0)$. The mass of the lowest lying 
state is then
$
m = \lim_{t\to \infty} \ln \frac{C^{(1)} (t)}{C^{(1)} (t-1)}
$.

Caveats of the mean-field are the gauge dependence of the background, 
the lack of guarantee that the expansion converges and 
the appearance of fake phase transitions.
Concerning the gauge dependence, physical observables
are found to be independent on the gauge fixing for the class of gauges
considered in Refs.~\citen{Irges:2009bi,Irges:2012ih}.
The mean-field corrections come multiplied by powers of $1/d$ (where $d$
is the number of dimensions) and therefore convergence is expected
to become better as the number of dimensions increases. In order to check
for the possibility of mean-field artifacts it is crucial to perform
Monte Carlo simulations. As we will see in Sec.~\ref{sec:mean-field-results}
and~\ref{sec:npghu-mf-results}, it turns out that the mean-field
in five dimensions captures the qualitative properties (like
dimensional reduction or spontaneous symmetry breaking) of the system.

\subsubsection{Mean-field on the torus}
\label{sec:mean-field-torus}

We present a quick way to determine the mean-field background.
The starting point are the link averages. In the case of $\SU(2)$ the link
average is defined by (see for example Ref.~\citen{Knechtli:1999tw})
\begin{equation}\label{e:SU2meanlink}
\langle U \rangle = \frac{
\int_{\SU(2)} \mathrm{d}U\, U\, {\rm e}^{\frac{1}{2}\mathrm{tr}\,(UB^\dagger)}}{
\int_{\SU(2)} \mathrm{d}U\, {\rm e}^{\frac{1}{2}\mathrm{tr}\,(UB^\dagger)}} 
= V \frac{I_2(b)}{I_1(b)} \,,
\end{equation}
where $B=bV$, $b\in\mathbb{R}$ and $V\in\SU(2)$.
The matrix $B$ originates from the gauge action and is equal, up to
a factor, to the sum of the staples.
In the case of $\U(1)$ the link average gives (cf. Ref.~\citen{Irges:2012ih})
\begin{equation}\label{e:U1meanlink} 
\langle U \rangle = \frac{
\int_{U(1)} \mathrm{d}U\, U\, {\rm e}^{\mathrm{Re}\,(UB^*)}}{
\int_{U(1)} \mathrm{d}U\, {\rm e}^{\mathrm{Re}\,(UB^*)}}
= V \frac{I_1(b)}{I_0(b)} \,, 
\end{equation}
where $B=bV$, $b\in\mathbb{R}$ and $V\in\U(1)$.

Consider the case of $\SU(2)$ on a $d$ dimensional torus.
The action is a generalization of Eq.~(\ref{e:torusaction}) to $d$ dimensions
where we set $\beta_4=\beta_5$.
The mean-field background is parametrized by the mean-link 
$U=u\times\mathbf{1}_2$.
The consistency condition that the link average Eq.~(\ref{e:SU2meanlink})
is equal to the mean-link, yields the relation
\begin{equation}\label{e:SU2mflinktorus}
b = 2(d-1)\,\beta\,u^3 \,,\quad u = \frac{I_2(b)}{I_1(b)} \,.
\end{equation}
It can be easily solved by numerical iteration. 

The generalization of Eq.~(\ref{e:SU2mflinktorus})
to the anisotropic gauge action is given in Ref.~\citen{Irges:2009bi},
where the five-dimensional $\SU(2)$ gauge theory is considered.
There explicit formulae for the calculation of the scalar and gauge boson
masses as well as for the static potential including fluctuations at leading
order are presented.

\subsubsection{Mean-field on the orbifold}
\label{sec:mean-field-orbifold}

On the orbifold translation invariance is broken. The mean-links are set to
\begin{eqnarray}
 U_\mu(n) & = & u(n_5)\,\times\,\mathbf{1}_2\,, \qquad\qquad\quad
 n_5\,=\,0,\ldots,N_5 \nonumber \\
 U_5(n)   & = & u(n_5+1/2)\,\times\,\mathbf{1}_2\,, \qquad 
 n_5\,=\,0,\ldots,N_5-1 \nonumber
\end{eqnarray}
Equating each link to its link average yields
a system of equations~\cite{Knechtli:2005dw,Irges:2012ih} which can
be solved by numerical iteration.
The link averages for the $\U(1)$ boundary links are computed using
Eq.~\eqref{e:U1meanlink} and for the $\SU(2)$ links using Eq.~\eqref{e:SU2meanlink}.


On the orbifold the calculation of the inverse propagator in 
Eq.~(\ref{e:mfinvprop}) is complicated by the breaking of translational 
invariance. This calculation is presented in Ref.~\citen{Irges:2012ih}.
There explicit formulae for the calculation of the scalar and gauge boson
masses as well as for the static potential including fluctuations at leading
order are also presented.

\subsection{Other models}
\label{sec:other-models}

In this section we present other models where five-dimensional gauge theories
have been studied. The first is a formulation of these theories with a warped
metric to study gauge field localization. The second is a
Lifshitz-type anisotropic formulation to study the possibility to take
a continuum limit.

Motivated by the Randall-Sundrum model~\cite{Randall:1999ee,Randall:1999vf} to 
localize gravity to four dimensions, in Ref.~\citen{Kenway:2015ofa} 
five-dimensional gauge theories have 
been considered on a warped background. The action is
\begin{eqnarray}\label{e:warpaction}
S_W^{AdS_5} & = & \sum_{n_{\mu}} \sum_{n_{5}=0}^{N_{5}-1} {\left[ 
\frac{\beta_{4}}{N_c}  \sum_{\mu<\nu}
{\real\tr\left\{1 - U_{\mu\nu}(n_{\mu},n_{5}) \right\}} + \right.} \nonumber \\
& & {\left. \frac{\beta_{5}}{N_c} \sum_\mu
{\real\tr\left\{ 1 - f(n_5)~U_{\mu5}(n_{\mu},n_{5}) \right\}} \right]}~,
\end{eqnarray}
where $f(n_5)=\exp(-2kn_5)$ is called the warp factor. So far 
Eq.~\eqref{e:warpaction} has 
been studied using the mean-field method only. The unconstrained variables 
$V$ in Eq.~\eqref{e:effaction} can be rescaled to absorb the warp factor in
$S_W^{AdS_5}(V)$. A rescaling of $H$ is then performed to keep the product
$HV$ unchanged. The resulting equations for the mean-field background 
with warping are similar to the ones 
discussed in Sec.~\ref{sec:mean-field-orbifold} for the orbifold geometry.
In Ref.~\citen{Kenway:2015ofa} the phase diagram and the static potential
have been studied for $N_c=2$. In particular fits to the static potential have been
performed and a Yukawa mass could be extracted. Close to the phase transition 
between the deconfined and the layered phase the shape of the static potential
is consistent with a nonzero four-dimensional Yukawa mass, hinting at the 
existence of a Higgs phase.
Earlier lattice studies of warped models in the context of gauge field
localization were carried out in Ref.~\citen{Laine:2002rh}.

In Ref.~\citen{Kanazawa:2014fla} $\SU(N_c)$ Lifshitz-type anisotropic
gauge theories proposed by Ho\v{r}ava were discretized on the lattice.
The lattice Ho\v{r}ava--Lifshitz action in $D+1$ dimensions is given by
\begin{equation}\label{e:lifshitzaction}
S = \frac{\beta_e}{2N_c} \sum_{n}\sum_{i=1}^D
{\real\tr\left\{1 - U_{0i}(n) \right\}} +
\frac{\beta_g}{2N_c} \sum_{n}\sum_{j=1}^D
{\real\tr\left\{1 - \prod_{\substack{i=1\\i\neq j}}^D T_{ij}(n) \right\}} \,,
\end{equation}
where $T_{ij}$ are twisted $2\times1$ Wilson loops in the $(i,j)$ plane,
see Ref.~\citen{Kanazawa:2014fla}. The ``time'' direction is idicated by
the subscript $0$ and $i,j=1,\ldots,D$ are the spatial directions.
The peculiarity of Eq.~\eqref{e:lifshitzaction} is that
the lattice spacing $b$ in the temporal direction has mass dimension $-2$ 
and the lattice spacing $a$ in the spacial directions has dimension $-1$.
The renormalization group equation at one-loop shows that the theory is
asymptotically free and the continuum limit is achieved at
$\beta_e,\beta_g\to\infty$~\cite{Kanazawa:2014fla}. First lattice
simulations of the $\SU(3)$ Lifshitz-type theory
present some evidence for this. There is no first order phase
transition visible in the action density. The expectation values of spatial
Wilson loops are found to be zero within errors consistently with the
absence of a term with spatial plaquettes $U_{ij}$ in 
Eq.~\eqref{e:lifshitzaction}.

\section{Pure gauge theory with compact extra dimensions}
\label{sec:pure-gauge-theory}

In the following we summarize results from several numerical studies of the simplest
non-Abelian extra-dimensional theory on the lattice, namely the $\SU(2)$
Yang--Mills theory on a five-dimensional torus with anisotropic
lattice spacings, $a_4 \ne a_5$~\cite{Ejiri:2000fc,Ejiri:2002ww,deForcrand:2010be,Knechtli:2011gq,Farakos:2010ie,DelDebbio:2012mr}.
Some lattice details are introduced in Sec.~\ref{sec:torus-geometry}.
Historically, this model with $a_4=a_5$ was studied in
the late seventies~\cite{Creutz:1979dw}.
It was recently extended
to higher dimensions $d>5$ in order to test
mean-field predictions~\cite{Irges:2015uta}.
This field of research has proven to be of great interest for a long time.

\subsection{Phase diagram}
\label{sec:phase-diagram}

The phase diagram of the lattice model has a rich structure that can
be identified by looking at the Polyakov loop expectation values along the fifth
dimension as a function of the bare parameters.
Three regions can be highlighted.
The location of these regions in the coupling space is summarized in Fig.~\ref{fig:phase-diagram}.
\begin{figure}[!ht]
\centerline{\includegraphics[scale=0.25, angle = 270]{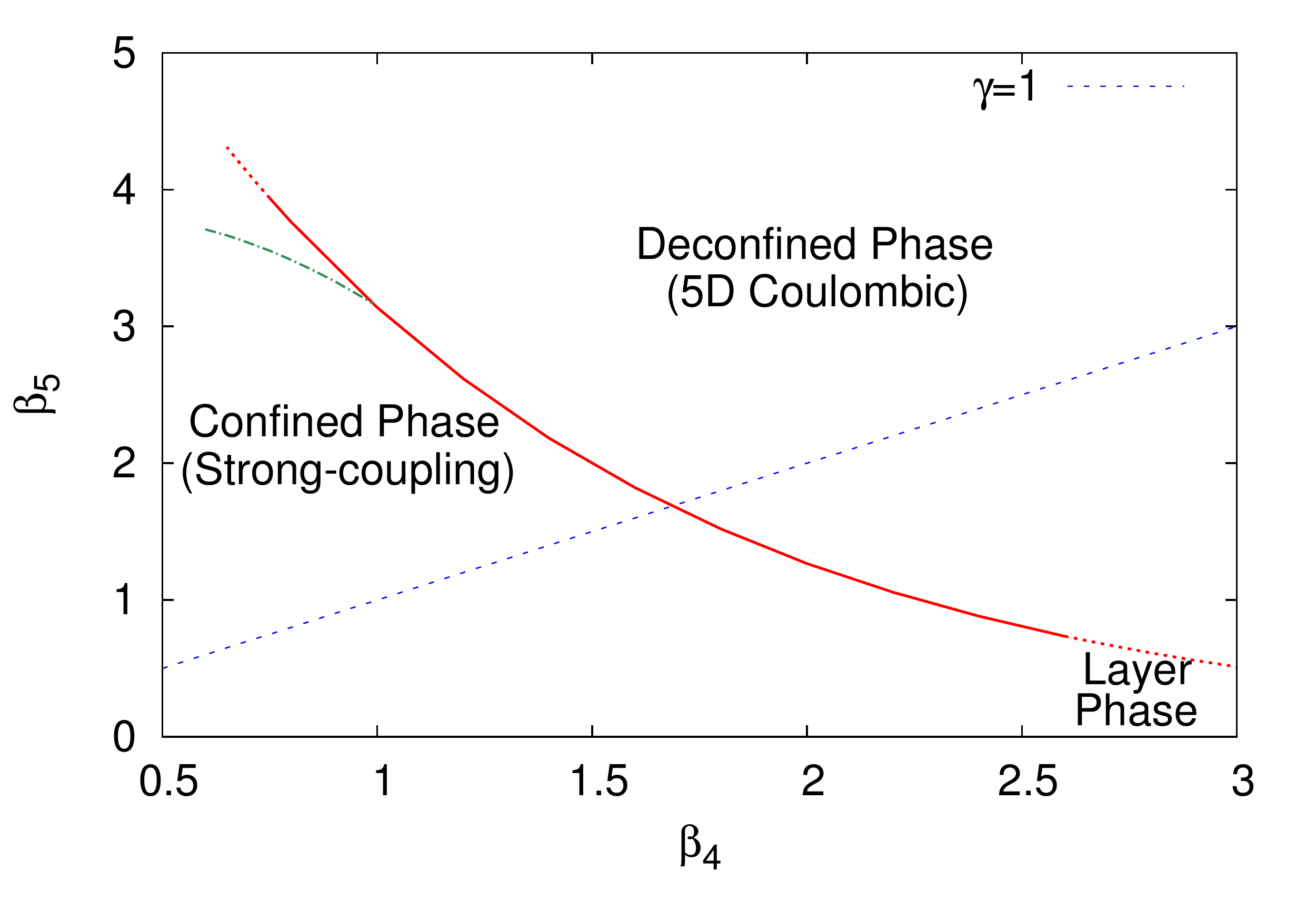}}
\caption{A sketch of the phase diagram of the anisotropic $\SU(2)$
  Yang--Mills model, taken from Ref.~\citen{DelDebbio:2013rka}.\label{fig:phase-diagram-cartoon}}
\end{figure}
When all the five dimensions are large, the system undergoes a first
order quantum (bulk) phase transition~\cite{Creutz:1979dw}  separating a confined phase from
a Coulomb phase (red solid line in Fig.~\ref{fig:phase-diagram-cartoon}).

When the fifth dimension is compactified at $\gamma>1$ where
$L_5 \ll L$, this bulk phase transition becomes a second order phase
transition separating a confined phase from a dimensionally reduced
phase with a non-zero four-dimensional string tension and a non-zero
scalar mass~\cite{deForcrand:2010be,DelDebbio:2012mr}.
This phase transition line changes with $N_5$ and corresponds to a
fixed critical value of the compactification radius $2\pi R=N_5a_5$.
It is indicated in Fig.~\ref{fig:phase-diagram-cartoon} by a green
dashed-dotted line.
The region with the desired separation of scales to describe a
four-dimensional gauge theory with a light scalar lies just above this
phase transition.

At $\gamma<1$, where $a_5 > a_4$, a different region called
\emph{Layer phase}~\cite{Knechtli:2011gq,Farakos:2010ie} appears as
predicted by mean-field
calculations~\cite{Irges:2009bi,Irges:2009qp}.
In this region, the planes transverse to the fifth dimension behave
as if they were decoupled from each other, indicating that degrees of
freedom are localized on the four-dimensional slices.
Mean-field predictions hint at a second order phase transition in this
region (dotted red line in Fig.~\ref{fig:phase-diagram-cartoon}), but Monte
Carlo simulations have not been able to conclusively determine this
feature~\cite{DelDebbio:2013rka}.
A similar localization mechanism was first observed for a $\U(1)$
Abelian gauge
theory~\cite{Dimopoulos:2006qz,Farakos:2008iw,Farakos:2008se} with
mean-field and Monte Carlo methods: the potential along $d=4$
hyperplanes is Coulombic, why it is confining in the bulk, constraining
fields to fluctuate only in four-dimensions.
Another idea, that achieves localization, relies on topological
defects, like a domain wall, and was demonstrated in Ref.~\citen{Dvali:1996bg}:
a confining potential in the bulk becomes deconfining on the defect,
where a theory with a reduced gauge symmetry group is realized.
An exploratory lattice study was carried over in Ref.~\citen{Laine:2004ji}.

To conclude, the phase diagram of the $d=5$ $\SU(2)$ Yang--Mills
theory on the lattice does not contain a second order phase
transition or a critical point where a five-dimensional continuum
theory can be defined non-perturbatively (contrary to expectations
from the $\epsilon$-expansion~\cite{Gies:2003ic,Morris:2004mg}).
However, regions of parameters where the theory dynamically reduces
from five to four dimensions exist at $\gamma>1$ and $\gamma<1$.

This phase diagram study has been extended to the larger gauge group
of $\SU(3)$ in Ref.~\citen{Beard:1997ic} and Ref.~\citen{Itou:2014iya}.
The reader should refer to those references for further details.

\subsection{Spectrum and static potential}
\label{sec:spectrum}

The nature of the dimensionally reduced theory can be
checked with non-perturbative calculations of the spectrum and of the
static potential.

At $\gamma>1$, the system is characterized by three energy scales
$\LambdaUV \approx \frac{1}{a_4}$, $\LambdaR \approx \frac{1}{N_5a_5}$
and $m_5$, pictorially shown in Fig.~\ref{fig:scale-sep}.
They were calculated for various values of the bare parameters of the
lattice model $\beta_4$, $\beta_5$ and $N_5$ (in the large-$N_4$
limit)~\cite{deForcrand:2010be,DelDebbio:2012mr} in units of the
four-dimensional lattice spacing, before rescaling them in units
of the four-dimensional string tension.
In other words, the lattice model can now be described in terms of what a
four-dimensional observer would measure: the string tension
$\sqrt{\sigma}$ is the inverse of the characteristic length in $d=4$.
Clearly this information is sufficient to check if the picture of
Fig.~\ref{fig:scale-sep} holds non-perturbatively.

Technically, the measurement of the four-dimensional string tension in units of the
lattice spacing,
$a_4\sqrt{\sigma}$, is done utilizing temporal correlation functions of spatial
Polyakov loops (averaged over the fifth dimension).
 The measurement of the scalar mass, again in units of the lattice
 spacing, $a_4m_5$, is instead performed using correlators 
of different types of four-dimensional scalar interpolating operators.
For example, a scalar operator from the four-dimensional point of view
is the Polyakov loop wrapping around the fifth dimension.
Another scalar operator in a four-dimensional gauge theory is a
glueball operator (a rotationally invariant combination of closed Wilson paths in
three dimensions).
They both carry scalar quantum numbers but the first is intrinsically
related to the extra-dimensional nature of the system.
Using both types of scalar operators is useful in disentangling
extra-dimensional contributions to the scalar spectrum.
A careful study of the scalar spectrum for a specific set of lattice
parameters is reported in Ref.~\citen{DelDebbio:2012mr} and the validity
of Eq.~\eqref{eq:one-loop-mass} in the non-perturbative regime is confirmed.

At $\gamma<1$, in the layer phase, the static potential was studied
numerically~\cite{Knechtli:2011gq} to check the mean-field
prediction~\cite{Irges:2009qp} that $d=4$ hyperplanes decouple, giving
rise to an effective dimensionally reduced $\SU(2)$ Yang--Mills theory
plus an adjoint scalar.
The static potential along the hyperplanes orthogonal to the fifth
dimension is compatible with a four-dimensional
description~\cite{Knechtli:2011gq} along the bulk phase transition
line, but still in the confined phase ($\beta_5<\beta_5^{\textrm{crit}}$). 
Moreover, the temporal Polyakov loops, in the phase where they have a
non-zero expectation value, fluctuate independently
depending on which $d=4$ hyperplane they originate from: at each $x_5$
coordinate along the extra dimension they have different central values.
Both these observations support the notion that dimensional reduction
happens at $\gamma<1$ due to a localization mechanism along the $d=4$
hyperplanes.

\subsection{Dimensional reduction and continuum limit}
\label{sec:dimens-reduct-cont}

\begin{figure}[!ht]
\centerline{\includegraphics[width=0.55\linewidth]{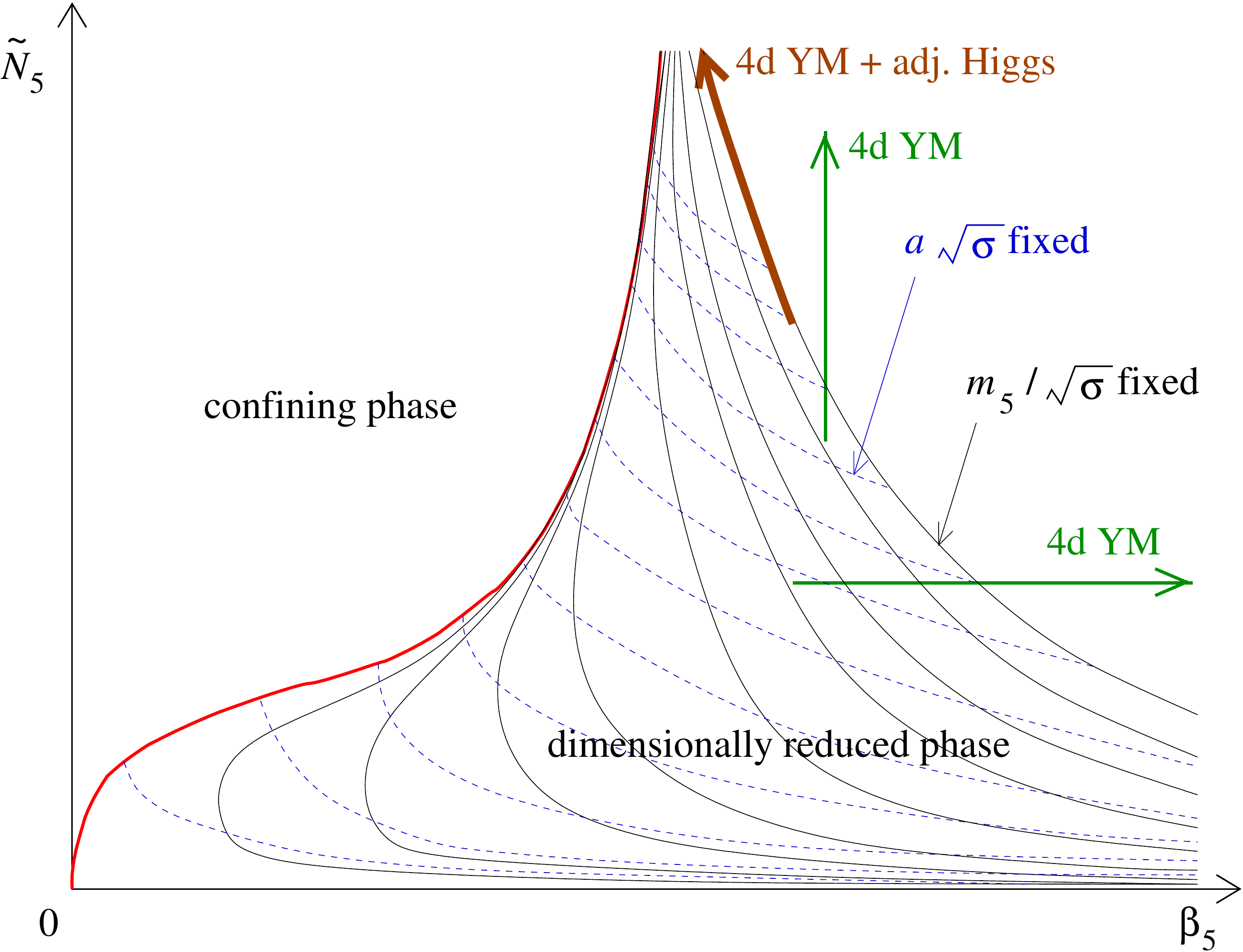}}
\caption{A sketch of the lines of constant physics for the anisotropic $\SU(2)$
  Yang--Mills model, taken from Ref.~\citen{deForcrand:2010be}.\label{fig:phase-diagram}}
\end{figure}

The non-perturbative studies of the scalar mass and of the static
potential for the anisotropic $\SU(2)$
  Yang--Mills model are useful to determine
the effective dimensionally-reduced description of the theory in
different regions of the phase diagram, and the corresponding lines of
constant physics.
Without lattice calculations, these lines can only be guessed with
perturbative predictions and may miss important non-perturbative contributions~\cite{Rinaldi:aa}.

The region at $\gamma>1$ has been studied with great details and
accuracy by different groups~\cite{deForcrand:2010be,DelDebbio:2012mr,Knechtli:2011gq}.
A sketch summarizing the results, taken from
Ref.~\citen{deForcrand:2010be}, is reported in Fig.~\ref{fig:phase-diagram}.
The figure shows a phase diagram in the ($\beta_5$,$\tilde{N}_5$)
plane (cfr. Sec.~\ref{sec:torus-geometry}) where a phase transition line of second
order separates the confined phase from the dimensionally reduced
phase where $L_5$ is small, as described in Sec.~\ref{sec:phase-diagram}.

The dashed lines on the right of the phase transition describe the
perturbative behavior of a constant four-dimensional string tension
and were checked
non-perturbatively~\cite{deForcrand:2010be,DelDebbio:2012mr}.
Similarly, the solid lines represent constant $m_5/\sqrt{\sigma}$,
again following a perturbative expectation that had been checked with
lattice numerical methods~\cite{deForcrand:2010be,DelDebbio:2012mr}.

A line of constant scalar mass $m_5$ in units of the four-dimensional theory
$\sqrt{\sigma}$ represents a Yang--Mills theory with an adjoint scalar
field.
Along this line, the cutoff is removed by going to larger $\tilde{N}_5$
values: the separation between the cutoff and the compactification
scale increases.
The scalar particle can be light or heavy depending on the size of the
extra dimension, with the lightest one being closest to the phase
transition line (where the correlation length $1/m_5$ diverges).
Any direction but the one along a constant $m_5/\sqrt{\sigma}$ line
results in a pure $d=4$ Yang--Mills theory when the cutoff is removed.
This can be easily seen because lines with heavier
$m_5/\sqrt{\sigma}$ are crossed in doing so: the scalar field
decouples and can be integrated out.
This is consistent with the na\"ive non-renormalizability of a
five-dimensional theory: removing the cutoff always yield a
four-dimensional Yang--Mills theory in the dimensionally-reduced phase.

The direction at constant $\tilde{N}_5$ was explicitly checked
non-perturbatively in Ref.~\citen{DelDebbio:2012mr}: the lightest
scalar of the theory becomes a glueball state as $\beta_5$ is
increased and the scalar coming from compactification is pushed to
higher energies.
A striking feature of this phase diagram, which is perhaps unexpected,
is that the theory is also a four-dimensional Yang--Mills theory along
the direction at fixed $\beta_5$ (as long as the phase transition is avoided.)
This direction corresponds to increasingly large compactification radius in
units of the four-dimensional cutoff.
The theory we end up with does not know about its extra-dimensional
origin, even when the extra dimension is large: this approach to
dimensional reduction was already suggested in the framework of
D-theories~\cite{Chandrasekharan:1996ih,Beard:1997ic,Brower:1997ha}.

\subsection{Mean-field results on the torus}
\label{sec:mean-field-results}

The investigation of the lattice phase diagram with Monte Carlo
methods was inspired by mean-field calculations described in 
Sec.~\ref{sec:mean-field-approach}.
The mean-field expansion has been applied to the
anisotropic $\SU(2)$ gauge theory on a periodic lattice in five dimensions in 
Refs.~\citen{Irges:2009bi,Irges:2009qp}.
The extension of the calculation to $\SU(N_c)$ was done in
Ref.~\citen{Irges:2012tu}.

The mean-field phase diagram has three phases: 
a confined phase, where the background vanishes everywhere; 
a deconfined phase, where the background is non-zero everywhere; and
a layered phase, where the background vanishes along the fifth dimension and is
nonzero elsewhere.
In Ref.~\citen{Irges:2009bi}, by analyzing the
shape of the static potential, two regions were identified, where
dimensional reduction occurs.
One region is for the anisotropy parameter
in Eq.~(\ref{e:torusaction}) $\gamma\gg1$ and it corresponds to a compact
extra dimension.
The other region is for
$\gamma<1$ close to the phase transition into the layered phase.
For sufficiently small $\gamma<1$
the phase transition, as seen by the mean-field, turns from first to second 
order.
When it is of second order a continuum limit can be taken.
In this case, dimensional reduction by localization on the four-dimensional
slices orthogonal to the extra dimension was shown in Ref.~\citen{Irges:2009qp}.
Although Monte Carlo studies did not confirm the existence of
a second order phase transition, they show hints for dimensional
reduction through localization at $\gamma<1$ \cite{Knechtli:2011gq}.
%
%

The mean-field expansion is expected to work better when the number of
dimensions $d$ is increased. 
On the isotropic torus the critical value $\beta_c^{\textrm{MF}}$ 
separates the confined from the Coulomb phase.
An interesting observation was made in Ref.~\citen{Irges:2015uta}.
The combination $(d-1)\,\beta_c^{\textrm{MF}}$ which fulfills 
Eq.~(\ref{e:SU2mflinktorus}) is independent on the number of dimensions $d$. 
This leads to the relation 
\begin{equation}\label{e:betacMF}
(d-1)\beta_c^{\textrm{MF}}\simeq6.704840 \,.
\end{equation}
This relation is compared to the location of the phase transition 
from Monte Carlo simulations in $d=5,6,7,8$ and shows a nice agreement.

\section{The Hosotani mechanism on the lattice}
\label{sec:ferm-hosot-mech}

The Hosotani mechanism~\cite{Hosotani:1983vn,Hosotani:1983xw} has been
established only in perturbation
theory, where an effective potential $\Veff(\AB)$ can be calculated
for the Aharonov-Bohm phase $\AB$, whose fluctuations correspond to
the Higgs mass, and remains free of divergences at one loop.
This spontaneous symmetry breaking mechanism that gives rise to the
Higgs and vector boson masses in the extra-dimensional
formalism was confirmed non-perturbatively with lattice field theory
methods in Ref.~\citen{Cossu:2013ora}.

Different realizations of gauge symmetry breaking described by the
Hosotani mechanism, correspond to different minima in
$\Veff(\AB)$.
The minima of the effective potential are located at different values
of $\AB$ which are probed by looking at the expectation value of the
Polyakov loop in the compact direction $y$:
\begin{equation}
  \label{eq:wilson-line-y}
  W(x) \, = \, P \exp \left( ig \int_{0}^{2\pi  R} dy \, A_y(x,y) \right) \comma
\end{equation}
where $g$ is the gauge coupling constant and
$A_y(x,y)$ the gauge potential in the compact direction with radius $R$.
The eigenvalues of $W(x)$, denoted by $\{e^{i\theta_{1}},e^{i\theta_{3}},e^{i\theta_{3}}\}$,
are the elements of the Aharonov-Bohm phase in the compact dimension
and represent the dynamical degrees of freedom which will
become the longitudinal components of the vector boson fields and of
the Higgs field in a complete Gauge-Higgs Unification scenario.

Ref.~\refcite{Cossu:2013ora}
studied the phase diagram of a ($3+1$)-dimensional $\SU(3)$ lattice
gauge theory with fermions in the adjoint (ad) and in the fundamental (fd)
representation with a compact direction.
The Hosotani mechanism works generically with any space-time
dimensionality, and this lattice theory is simpler
to study than its five-dimensional counterpart (due to the inherent
difficulty of formulating discretized versions of the fermionic
algebra in extra dimensions).
The minima of $\Veff(\AB)$ are investigated non-perturbatively in this
setup.

The lattice version of Eq.~\eqref{eq:wilson-line-y}, denoted by $P_3$, and its adjoint
counterpart, $P_8$, are computed for different values of gauge coupling
$\beta$ and fermion masses $m_{\textrm{fd}},\ m_{\textrm{ad}}$, as well
as different boundary conditions $\alpha_{\textrm{fd}}$ for
the fundamental fermions in the compact direction.
The distribution of eigenvalues of $P_3$ is compared to the
values of $\AB=(\theta_1,\theta_2,\theta_3)$ at the minima of
the perturbative one-loop effective potential.
%
%

Let us focus on the case where the $\SU(3)$ gauge theory is coupled to
two massive adjoint fermions with periodic boundary conditions on the
toroidal lattice~\cite{Cossu:2009sq}.
By looking at the expectation value of $P_3$, one can identify four different
phases at fixed $m_{\textrm{ad}}$ as the extra dimension size
decreases by increasing $\beta$: they are called, in order, X, A, B
and C.
Perturbatively, different phases are related to different configurations
of the minima of $\Veff(\AB)$ and they appear as the product
$m_{\textrm{ad}}R$ gets smaller.

The \emph{X phase} corresponds to a confined $\SU(3)$ symmetric phase which
disappears when the extra dimension starts to shrink.
The \emph{A phase} corresponds to a degenerate triplet of minima for the
effective potential: $P_3$ takes the values of the cubic root of unity
with equal probability and the
$\theta_i$ eigenvalues are always degenerate for each case (modulo the
Haar measure).
This phase is deconfined, but still $\SU(3)$ symmetric.
If one keeps driving $R$ to zero, the \emph{B phase} appears (called
\emph{split} in the original study~\cite{Cossu:2009sq}): in terms of
the Hosotani mechanism, this phase
corresponds to the $\SU(3)$ gauge symmetry being broken to $\SU(2)
\times \U(1)$ and this can be seen by investigating the eigenvalues
$\theta_i$.
Again there is a triplet of minima where two elements of $\AB$ are
degenerate while the third is different. 
At last, the \emph{C phase}, originally called
\emph{reconfined}~\cite{Cossu:2009sq}, has $\langle P_3 \rangle = 0$, but it corresponds
to a $\U(1) \times \U(1)$ gauge symmetry, where the eigenvalues
$\theta_i$ take maximally displaced values: $(0,\frac{2}{3}\pi,-\frac{2}{3}\pi)$.
In the \emph{B} and \emph{C phases}, the masses of the adjoint scalars
are calculated by looking at the fluctuations around the minima of the
potential and they agree with the perturbative predictions.

The correspondence between the perturbative Hosotani mechanism
and the non-perturbative phases on the lattice was also checked
for the case of a $\SU(3)$ gauge theory with four fundamental
fermions by changing their temporal boundary conditions.
However, these studies have not attempted to investigate if this
correspondence survives in the continuum limit of the lattice theory.

It is interesting to note that the lattice phase diagrams for the two
theories described above were originally investigated for reasons that are not
related to the Hosotani mechanism.
The theory with adjoint fermions was studied in
Ref.~\citen{Cossu:2009sq} to study volume independence in the orientifold
planar equivalence~\cite{Kovtun:2007py}.
On the other hand, different boundary conditions for fundamental
fermions were studied in the context of finite density QCD with
imaginary chemical potential~\cite{deForcrand:2002hgr}.

A first study of a four-dimensional theory, where spontaneous
symmetry breaking is induced by the Hosotani mechanism in five
dimensions, was performed in Ref.~\citen{Akerlund:2015poa}.
An external potential
$h_{\textrm{fd}}\real (\Tr P_5) + h_{\textrm{ad}} | \Tr P_5 |^2$, based on
the Polyakov loop in the extra dimension $P_5$, is used to mimic the
presence of adjoint fermions and break the gauge symmetry. 
In the vacuum where the gauge symmetry group
is $\U(1) \times \U(1)$ (cfr. the C phase above), a topological
excitation which is gauge invariant only under a $\U(1)$ group, namely
an Abelian flux, is found to be stable, while it would immediately
disappear when the full $\SU(3)$ gauge symmetry is restored by
switching off the external potential.

\section{Non-perturbative gauge-Higgs unification}
\label{sec:non-pert-gauge}

The phenomenological interest in orbifold theories relies on the possibility 
to break spontaneously the boundary gauge group $H$ (see Sec.~
\ref{sec:orbifold-geometry}) and doing so to reproduce 
the Higgs mechanism of the Standard Model.
Perturbative calculations of the effective potential for the extra-dimensional 
scalar indicate that spontaneous symmetry breaking cannot 
occur unless fermions are included, see Ref.~\citen{Kubo:2001zc}.
The perturbative limit of these theories is trivial and one wonders about the
situation in the non-perturbative regime.
The findings of the earlier Monte Carlo studies of the $\SU(2)$ pure gauge
theory on the orbifold~\cite{Irges:2006zf,Irges:2006hg} showed indeed evidence 
for a massive gauge boson.

\subsection{Mean-field results on the orbifold}
\label{sec:npghu-mf-results}

\begin{figure}[h]
\centerline{\includegraphics[height=4cm,width=6cm]{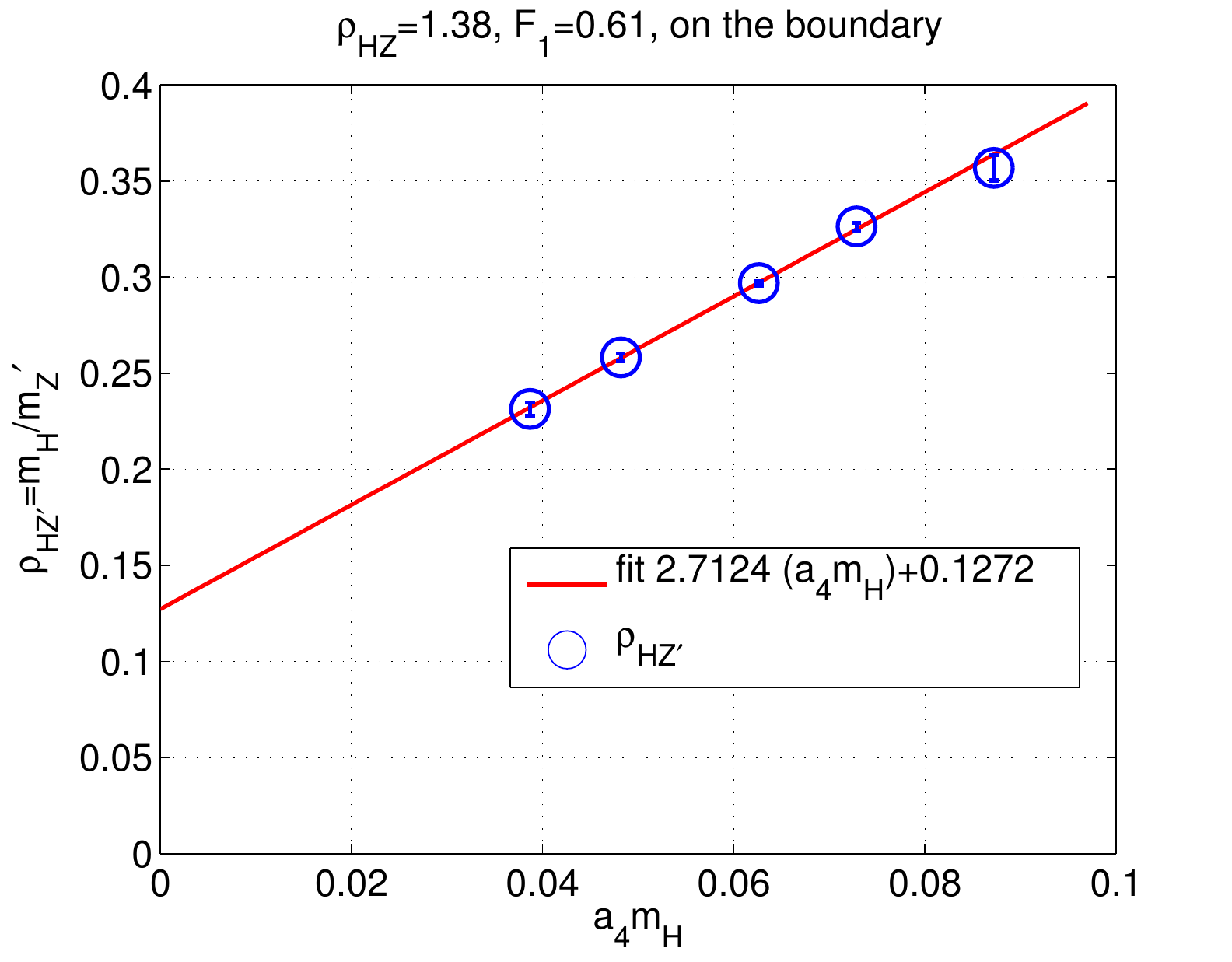}}
\caption{The ratio $\frac{m_{H}}{m_{Z'}}$
computed in mean-field on the orbifold. From Ref.~\citen{Irges:2012mp}.
\label{f_GHU_mf}}
\end{figure}

A first attempt to probe analytically five-dimensional gauge theories on the
orbifold away
from the trivial perturbative limit was made in Ref.~\citen{Irges:2007qq}.
Including the effects of a finite lattice spacing in the Coleman-Weinberg
computations~\cite{Antoniadis:2001cv} of the effective potential leads to 
the possibility of spontaneous symmetry breaking in the pure gauge theory,
contrary to the expectation based on perturbative calculations~
\cite{Kubo:2001zc} but in accordance with the Monte Carlo results~
\cite{Irges:2006zf,Irges:2006hg}.
A further verification of the existence
of spontaneous symmetry breaking was provided by the mean-field calculations
in Refs.~\citen{Irges:2012ih,Irges:2012mp}.
The scalar mass is extracted from the correlator of scalar Polyakov loops. 
The gauge boson mass is obtained by determining a four-dimensional
Yukawa mass from the static potential~\cite{Irges:2012ih}. 
It turns out to be non-zero on
the orbifold in the limit of an infinitely large boundary, 
which is the evidence for 
the spontaneous breaking of the boundary $\U(1)$ gauge symmetry.
The same mass extracted on the torus vanishes instead.

At sufficiently small $\gamma$ it is possible to construct lines of
constant physics in the deconfined phase, where the mean-field background 
is non-zero everywhere. The lines
are close to the phase boundary and in the regime where the phase transition
is of second order according to the mean-field.
In Ref.~\citen{Irges:2012mp} such a line is constructed where the Higgs to 
$Z$ boson ratio is kept fixed to $\rho \equiv \frac{m_{H}}{m_{Z}}= 1.38$ and 
the Higgs mass is kept fixed to $F_1=m_HR=0.61$, where $\pi R$ is the size
of the extra dimension.
From the boundary static potential it is possible to extract also a
higher Yukawa mass which corresponds to an excited $Z'$ boson state.
Fig.~\ref{f_GHU_mf} shows the continuum extrapolation of the ratio
$\frac{m_{H}}{m_{Z'}}$. For $m_Z=91.19\,\mathrm{GeV}$ the result is
a $Z'$ mass of $989\,\mathrm{GeV}$. We notice that the fact that 
the ratio shown Fig.~\ref{f_GHU_mf} can be extrapolated also implies the
finiteness of the Higgs mass.

\subsection{Monte Carlo results on the orbifold}
\label{sec:npghu-mc-results}

The latest Monte Carlo results, which are favorably pointing towards the 
suitability of this theory for describing the electro-weak sector of the 
Standard Model are reported in Ref.~\citen{Alberti:2015pha},
of which this section is a summary.

\subsubsection{Phase diagram}
\label{sec:npghu-phase-diagram-results}

\begin{figure}[h]
\centerline{\includegraphics[width=8cm]{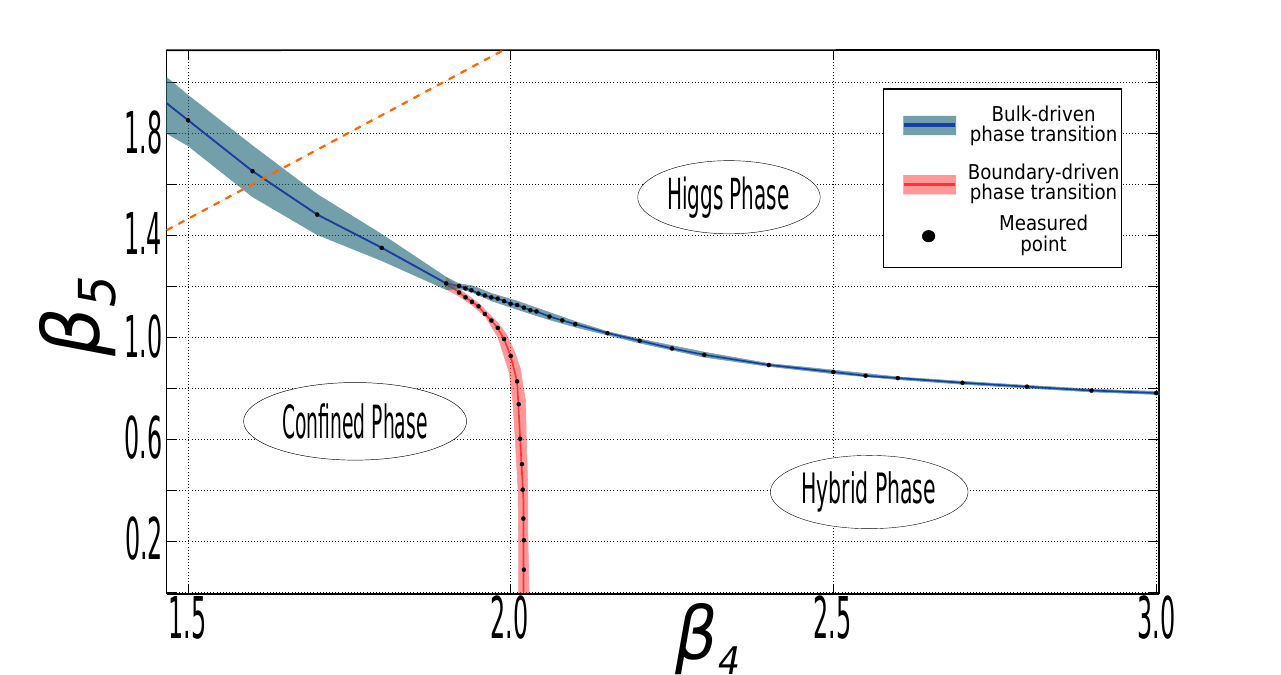}}
\caption{The phase diagram of the five-dimensional anisotropic orbifold 
($N_5=4$) in the region of the Higgs-hybrid phase transition. 
From Ref.~\citen{Alberti:2015pha}. \label{f_GHU_pd}}
\end{figure}

In Ref.~\citen{Alberti:2015pha}, a five-dimensional $\SU(2)$ theory was 
studied on an anisotropic lattice.
On the boundaries, where the "Standard Model-like" physics would be found, 
the gauge group reduces to $\U(1)$ due to the orbifold geometry.
The study of this model is intended as a proof of principle: the investigation 
of a larger model which would contain,
on the boundary, the entirety of the Higgs sector will follow once the theory 
has been shown to be viable.

The system has been found to exhibit three phases, characterized by the 
expectation value of the Polyakov loop: in the confined
(de-confined) phase the Polyakov loop exhibit zero (non-zero) expectation 
value in every direction. In this context, the de-confined phase is labelled
Higgs phase, because it is where the Higgs potential develops SSB, giving rise 
to non-zero gauge boson masses.
The third phase, which is characteristic only of the orbifold geometry, shows 
confined dynamics in the orbifold's bulk, and de-confined
dynamics on its boundaries; it is, therefore, called hybrid phase. 
Both phase transitions have been found to be first order for
all explored parameter values. The phase diagram in the $(\beta_4,\beta_5)$
plane for the orbifold with $N_5=4$ is shown in Fig.~\ref{f_GHU_pd}.
The Figure shows the region where the bare anisotropy $\gamma\le1$. The
phase transition line separating the confined and the hybrid phases
originates from the phase transition of the four-dimensional $\U(1)$ theory
on the orbifold boundaries.
Its appearance is crucial because it changes the properties of the
spectrum in the Higgs phase, as we describe below.
 
\subsubsection{Spectrum and the Higgs mechanism}
\label{sec:npghu-spectrum-results}

\begin{figure}[t]
\centerline{\includegraphics[width=7cm]{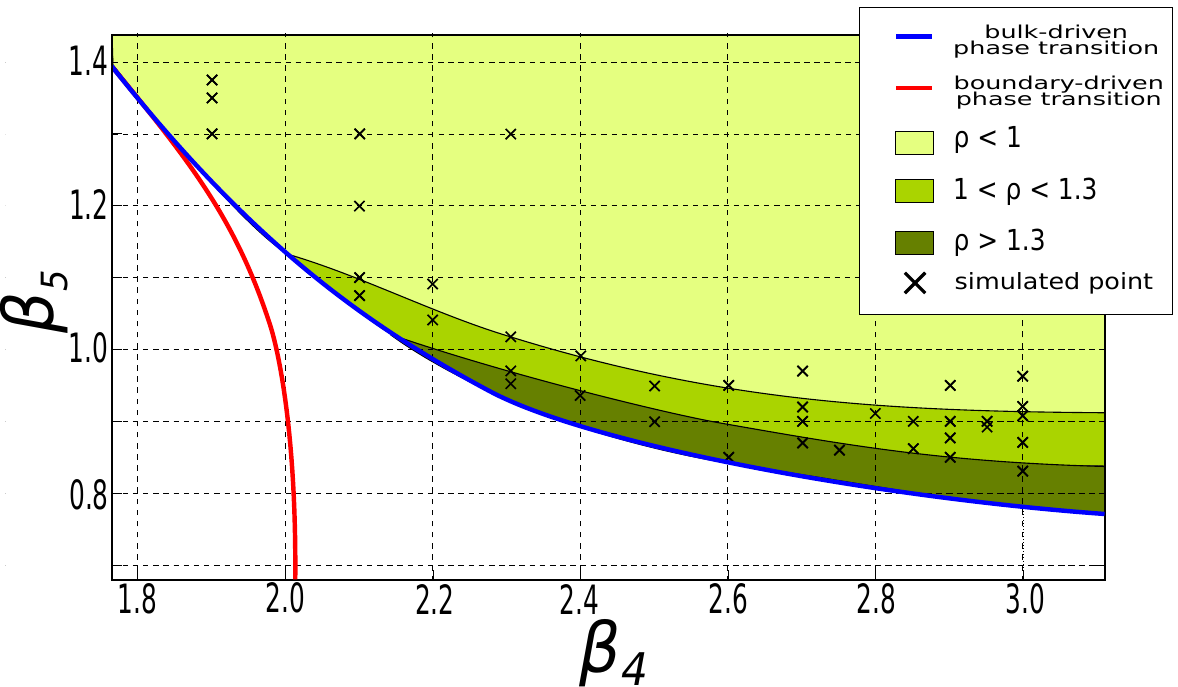}}
\caption{The ratio $\frac{m_{H}}{m_{Z}}$ on the orbifold. 
The different shadings of green correspond to $\rho < 1$ 
(lightest shade), $1 \leq \rho \leq 1.3$ (middle shade) and
$\rho > 1.3$ (darkest shade). From Ref.~\citen{Alberti:2015pha}.
\label{f_GHU_spectrum}}
\end{figure}

Although the Higgs mechanism is active everywhere in the Higgs phase, the mass 
measurements of the scalar and vector boson
(which we identify with the \emph{Z} boson, due to the quantum numbers of the 
operators we use)
suggest that only a smaller region of the parameter space is capable of 
reproducing the physics of the Standard Model.
Fig.~\ref{f_GHU_spectrum} shows the observed value of the ratio 
$\rho \equiv \frac{m_{H}}{m_{Z}}$ on top of the
theory's phase diagram: the region in proximity of the Higgs-hybrid phase 
transition is where $\rho$ approaches
the physical value of $ ~ 1.37$.

\subsubsection{Dimensional reduction by localization}
\label{sec:npghu-dim-red-results}

\begin{figure}[t]
\centerline{\includegraphics[width=11cm]{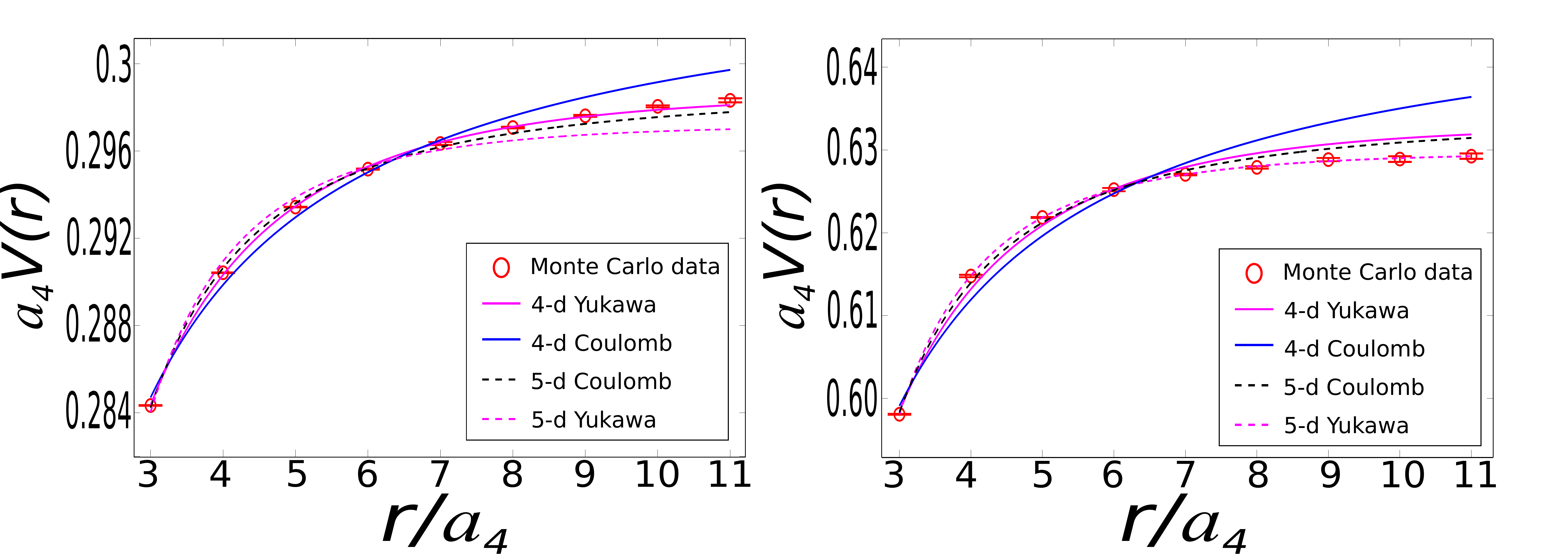}}
\caption{The static quark-antiquark potential on the orbifold measured 
on the boundary (left) and in the bulk (right) in proximity of the
Higgs-hybrid phase transition.
From Ref.~\citen{Alberti:2015pha}. \label{f_GHU_potential}}
\end{figure}

Exactly in the same region, where the masses of the Higgs and \emph{Z} boson
show the physical ratio,
dimensional reduction is observed.
As shown in Fig.~\ref{f_GHU_potential}, the static potential measured on the 
boundary (left panel) at $\left(\beta_{4}=2.1, \beta_{5}=1.075, N_{5}=4\right)$
is of four-dimensional type, whereas the same observable measured in the 
orbifold's bulk (right panel) is clearly five-dimensional.
Moreover, the mass of the \emph{Z} boson, which can be extracted from the fit 
to the four-dimensional Yukawa potential, has been found to be in perfect 
agreement with the mass measured in the same point through spectroscopic 
calculations.
Contrastingly, in the region further away from the phase transition, 
where the hierarchy of the masses is unphysical,
no hint of four-dimensional physics could be observed neither on the 
orbifold's boundaries nor in the bulk.

On the other side of the phase transition, in the hybrid phase, the static 
potential is found to be four dimensional on all
space-time layers. The $U(1)$ boundaries exhibit de-confined physics, 
consistently with the behaviour of a pure 4-d $U(1)$ theory at a matching 
value of the coupling $\beta=\frac{\beta_{4}}{2}$~\cite{Arnold:2002jk}. 
On the other hand, the $SU(2)$ bulk layers show confinement.
Although this phase is not of immediate interest in the study of the 
Higgs mechanism, as it does not possess SSB, a deeper
knowledge of what happens in it could prove a useful tool in understanding 
processes in the Higgs phase.

While all the results presented here have been obtained on $N_{5}=4$ lattices, 
measurements on $N_{5}>4$ have shown no
qualitative difference from those quoted, the only change being that the 
masses become lighter in lattice units and the
region of the phase diagram where the physical mass hierarchy is found become 
slightly larger.
This suggests, consistently with the idea of localization, that the extent of 
the extra dimension plays only a minor
role in the resulting physics.

\subsubsection{Effective theory and the Higgs mechanism}

Global symmetries are essential for the Higgs mechanism to take place in the 
non-perturbative gauge invariant formulation of the theory on the lattice. 
Elitzur's theorem dictates that any physical effect associated with the 
breaking of a local symmetry must originate from the spontaneous breaking of a 
non-trivial global symmetry~\cite{Elitzur:1975im}.

In Refs.~\citen{Irges:2013rya,Alberti:2015pha} it is shown that
the spontaneous breaking of the boundary gauge symmetry $H$ on the orbifold
is governed by the stick symmetry~\cite{Ishiyama:2009bk}
$\mathcal{S} = \mathcal{S}_{L} \cdot \mathcal{S}_{R}$
defined in Eqs.~(\ref{e:stickleft}) and (\ref{e:stickright}).
The order parameter is
is the vector Polyakov loop defined in Eq.~(\ref{e:gaugebosonorb}), which
is odd under the stick symmetry $\mathcal{S}$ and from which a
nonzero gauge boson mass can be determined.
The center symmetry is obtained by applying twice the stick symmetry.
The scalar and vector Polyakov loops on the orbifold are invariant under the 
center symmetry (unlike on the torus). As a result,
the spontaneous breaking of the stick symmetry does not trigger 
finite temperature phase transitions but a Higgs-mechanism like phase 
transition.

The stick symmetry appears to be valid only at finite lattice spacing
~\cite{Alberti:2015pha}.
In the perturbative continuum limit the symmetry disappears. This 
explains the absence of spontaneous symmetry breaking in pure gauge 
perturbative models of gauge-Higgs unification~\cite{Kubo:2001zc}.
In non-perturbative gauge-Higgs unification
the spontaneous breaking of the stick symmetry and the resulting Higgs 
mechanism make sense only as an effective theory.
The construction of lines of constant physics which describe 
the effective theory in a finite range of lattice spacings is work in
progress. If they can be constructed, they will be a proof of the
(non-perturbative) finiteness of the Higgs mass. In addition, they
will predict the energy of the excited states of the Higgs and gauge boson 
particles to be compared with experimental results.

\section{Perspectives}
\label{sec:conclusions}

Extra-dimensional models are actively being searched for at the
LHC.
Non-perturbative investigations using lattice field theory methods,
described in this review, have proven successful in testing several
scenarios relevant for gauge-Higgs unification.
Despite the long history of numerical simulations for
extra-dimensional models, there is still no complete description that
is directly applicable to phenomenology. 
Some of the main reasons for this are listed in the following.
They should be used as a reference guide for future works aimed at
improving our understanding of phenomenologically relevant
extra-dimensional models.

Lattice models with one or more extra dimensions are inherently more
expensive in terms of computing cost, and the same is true for larger
gauge groups which are needed to embed the electro-weak sector of the
Standard Model.
Moreover, it is technically challenging to define discretized versions
of the fermionic algebra when the number of space-time dimensions is
not even.
Furthermore, it seems that a non-trivial continuum limit of
extra-dimensional gauge theories is hard to achieve.
Even if such a limit does not exist, lines of constant physics can be 
constructed when lattice models are interpreted as effective field theory
descriptions with an ultraviolet cutoff.
There are investigations of alternative gauge theories where a continuum
limit in higher dimension might exist.
More sophisticated extra-dimensional models are based on a warped metric
and lattice studies consider also these models.

To conclude, we think that this is a very active field and will see
significant advances in the future.

\section*{Acknowledgments}

We thank N.~Irges and G.~Cossu for useful comments.
F.~K. scknowledges the support of the Deutsche Forschungsgemeinschaft (DFG) 
under contract {KN 947/1-2}.
E.~R. acknowledges the support of the DOE under contract {DE-AC52-07NA27344} (LLNL).

\bibliographystyle{ws-ijmpa}
\bibliography{master_review}

\begin{thebibliography}{10}
\expandafter\ifx\csname urlstyle\endcsname\relax
  \providecommand{\doi}[1]{doi:\discretionary{}{}{}#1}\else
  \providecommand{\doi}{doi:\discretionary{}{}{}\begingroup
  \urlstyle{rm}\Url}\fi

\bibitem{ATLAS:2012gk}
 ATLAS Collaboration Collaboration (G.~Aad {\em et~al.}), {\em Phys.Lett.} {\bf
  B716}, 1  (2012), \href{http://arxiv.org/abs/1207.7214}{{\ttfamily
  1207.7214}}.

\bibitem{CMS:2012gu}
 CMS Collaboration Collaboration (S.~Chatrchyan {\em et~al.}), {\em Phys.Lett.}
  {\bf B716}, 30  (2012), \href{http://arxiv.org/abs/1207.7235}{{\ttfamily
  1207.7235}}.

\bibitem{Englert:1964et}
F.~Englert and R.~Brout, {\em Phys.Rev.Lett.} {\bf 13}, 321  (1964).

\bibitem{Higgs:1964ia}
P.~W. Higgs, {\em Phys.Lett.} {\bf 12}, 132  (1964).

\bibitem{Higgs:1964pj}
P.~W. Higgs, {\em Phys.Rev.Lett.} {\bf 13}, 508  (1964).

\bibitem{CMS:2015dxe}
CMS Collaboration, { {Search for new physics in high mass diphoton events in
  proton-proton collisions at $\sqrt{s} = 13$ TeV}}, in {\em {ATLAS and CMS
  physics results from Run 2, CERN, 15 December}\/},  (2015).

\bibitem{ATLAS:2015hp}
ATLAS Collaboration, { {Search for resonances decaying to photon pairs in 3.2
  fb$^{-1}$ of $pp$ collisions at $\sqrt{s}$ = 13 TeV with the ATLAS
  detector}}, in {\em {ATLAS and CMS physics results from Run 2, CERN, 15
  December}\/},  (2015).

\bibitem{Kaluza:1921tu}
T.~Kaluza, {\em Sitzungsber.Preuss.Akad.Wiss.Berlin (Math.Phys.)} {\bf 1921},
  966  (1921).

\bibitem{Klein:1926tv}
O.~Klein, {\em Z.Phys.} {\bf 37}, 895  (1926).

\bibitem{Manton:1979kb}
N.~Manton, {\em Nucl.Phys.} {\bf B158},   141  (1979).

\bibitem{Hosotani:1983xw}
Y.~Hosotani, {\em Phys. Lett.} {\bf B126},   309  (1983).

\bibitem{Hatanaka:1998yp}
H.~Hatanaka, T.~Inami and C.~S. Lim, {\em Mod. Phys. Lett.} {\bf A13}, 2601
  (1998), \href{http://arxiv.org/abs/hep-th/9805067}{{\ttfamily
  hep-th/9805067}}.

\bibitem{Cheng:2002iz}
H.-C. Cheng, K.~T. Matchev and M.~Schmaltz, {\em Phys. Rev.} {\bf D66},
  036005  (2002), \href{http://arxiv.org/abs/hep-ph/0204342}{{\ttfamily
  hep-ph/0204342}}.

\bibitem{vonGersdorff:2002as}
G.~von Gersdorff, N.~Irges and M.~Quiros, {\em Nucl. Phys.} {\bf B635}, 127
  (2002), \href{http://arxiv.org/abs/hep-th/0204223}{{\ttfamily
  hep-th/0204223}}.

\bibitem{Hosotani:2005fk}
Y.~Hosotani  (2005), \href{http://arxiv.org/abs/hep-ph/0504272}{{\ttfamily
  hep-ph/0504272}}.

\bibitem{Hosotani:2007kn}
Y.~Hosotani, N.~Maru, K.~Takenaga and T.~Yamashita, {\em Prog. Theor. Phys.}
  {\bf 118}, 1053  (2007), \href{http://arxiv.org/abs/0709.2844}{{\ttfamily
  0709.2844}}.

\bibitem{DelDebbio:2008hb}
L.~Del~Debbio, E.~Kerrane and R.~Russo, {\em Phys. Rev.} {\bf D80},   025003
  (2009), \href{http://arxiv.org/abs/0812.3129}{{\ttfamily 0812.3129}}.

\bibitem{Elitzur:1975im}
S.~Elitzur, {\em Phys.Rev.} {\bf D12}, 3978  (1975).

\bibitem{Fu:1983ei}
Y.~Fu and H.~B. Nielsen, {\em Nucl.Phys.} {\bf B236},   167  (1984).

\bibitem{Ejiri:2000fc}
S.~Ejiri, J.~Kubo and M.~Murata, {\em Phys. Rev.} {\bf D62},   105025  (2000),
  \href{http://arxiv.org/abs/hep-ph/0006217}{{\ttfamily hep-ph/0006217}}.

\bibitem{Ejiri:2002ww}
S.~Ejiri, S.~Fujimoto and J.~Kubo, {\em Phys.Rev.} {\bf D66},   036002  (2002),
  \href{http://arxiv.org/abs/hep-lat/0204022}{{\ttfamily hep-lat/0204022}}.

\bibitem{Irges:2009bi}
N.~Irges and F.~Knechtli, {\em Nucl. Phys.} {\bf B822}, 1  (2009),
  \href{http://arxiv.org/abs/0905.2757}{{\ttfamily 0905.2757}}.

\bibitem{Farakos:2010ie}
K.~Farakos and S.~Vrentzos, {\em Nucl.Phys.} {\bf B862}, 633  (2012),
  \href{http://arxiv.org/abs/1007.4442}{{\ttfamily 1007.4442}}.

\bibitem{deForcrand:2010be}
P.~de~Forcrand, A.~Kurkela and M.~Panero, {\em JHEP} {\bf 06},   050  (2010),
  \href{http://arxiv.org/abs/1003.4643}{{\ttfamily 1003.4643}}.

\bibitem{Knechtli:2011gq}
F.~Knechtli, M.~Luz and A.~Rago, {\em Nucl.Phys.} {\bf B856}, 74  (2012),
  \href{http://arxiv.org/abs/1110.4210}{{\ttfamily 1110.4210}}.

\bibitem{DelDebbio:2012mr}
L.~Del~Debbio, A.~Hart and E.~Rinaldi, {\em JHEP} {\bf 1207},   178  (2012),
  \href{http://arxiv.org/abs/1203.2116}{{\ttfamily 1203.2116}}.

\bibitem{Ejiri:1998xd}
S.~Ejiri, Y.~Iwasaki and K.~Kanaya, {\em Phys. Rev.} {\bf D58},   094505
  (1998), \href{http://arxiv.org/abs/hep-lat/9806007}{{\ttfamily
  hep-lat/9806007}}.

\bibitem{Dienes:1998vg}
K.~R. Dienes, E.~Dudas and T.~Gherghetta, {\em Nucl. Phys.} {\bf B537}, 47
  (1999), \href{http://arxiv.org/abs/hep-ph/9806292}{{\ttfamily
  hep-ph/9806292}}.

\bibitem{Irges:2007qq}
N.~Irges, F.~Knechtli and M.~Luz, {\em JHEP} {\bf 0708},   028  (2007),
  \href{http://arxiv.org/abs/0706.3806}{{\ttfamily 0706.3806}}.

\bibitem{Irges:2004gy}
N.~Irges and F.~Knechtli, {\em Nucl. Phys.} {\bf B719}, 121  (2005),
  \href{http://arxiv.org/abs/hep-lat/0411018}{{\ttfamily hep-lat/0411018}}.

\bibitem{Knechtli:2005dw}
F.~Knechtli, B.~Bunk and N.~Irges, {\em PoS} {\bf LAT2005},   280  (2006),
  \href{http://arxiv.org/abs/hep-lat/0509071}{{\ttfamily hep-lat/0509071}}.

\bibitem{Hebecker:2001jb}
A.~Hebecker and J.~March-Russell, {\em Nucl.Phys.} {\bf B625}, 128  (2002),
  \href{http://arxiv.org/abs/hep-ph/0107039}{{\ttfamily hep-ph/0107039}}.

\bibitem{Irges:2006hg}
N.~Irges and F.~Knechtli, {\em Nucl. Phys.} {\bf B775}, 283  (2007),
  \href{http://arxiv.org/abs/hep-lat/0609045}{{\ttfamily hep-lat/0609045}}.

\bibitem{Ishiyama:2009bk}
K.~Ishiyama, M.~Murata, H.~So and K.~Takenaga, {\em Prog.Theor.Phys.} {\bf
  123}, 257  (2010), \href{http://arxiv.org/abs/0911.4555}{{\ttfamily
  0911.4555}}.

\bibitem{Irges:2013rya}
N.~Irges and F.~Knechtli, {\em JHEP} {\bf 1406},   070  (2014),
  \href{http://arxiv.org/abs/1312.3142}{{\ttfamily 1312.3142}}.

\bibitem{Drouffe:1983fv}
J.-M. Drouffe and J.-B. Zuber, {\em Phys.Rept.} {\bf 102},  ~1  (1983).

\bibitem{Ruhl:1982er}
W.~R{\"u}hl, {\em Z. Phys.} {\bf C18},   207  (1983).

\bibitem{Irges:2012ih}
N.~Irges, F.~Knechtli and K.~Yoneyama, {\em Nucl.Phys.} {\bf B865}, 541
  (2012), \href{http://arxiv.org/abs/1206.4907}{{\ttfamily 1206.4907}}.

\bibitem{Knechtli:1999tw}
F.~Knechtli, {The Static potential in the SU(2) Higgs model}, PhD thesis,
  Humboldt U., Berlin  (1999).

\bibitem{Randall:1999ee}
L.~Randall and R.~Sundrum, {\em Phys. Rev. Lett.} {\bf 83}, 3370  (1999),
  \href{http://arxiv.org/abs/hep-ph/9905221}{{\ttfamily hep-ph/9905221}}.

\bibitem{Randall:1999vf}
L.~Randall and R.~Sundrum, {\em Phys. Rev. Lett.} {\bf 83}, 4690  (1999),
  \href{http://arxiv.org/abs/hep-th/9906064}{{\ttfamily hep-th/9906064}}.

\bibitem{Kenway:2015ofa}
R.~D. Kenway and E.~Lambrou, { {Five-dimensional Gauge Theories in a warped
  background}}, in {\em {Proceedings, 33rd International Symposium on Lattice
  Field Theory (Lattice 2015)}\/},  (2015).
\newblock \href{http://arxiv.org/abs/1510.07601}{{\ttfamily 1510.07601}}.

\bibitem{Laine:2002rh}
M.~Laine, H.~B. Meyer, K.~Rummukainen and M.~Shaposhnikov, {\em JHEP} {\bf 01},
    068  (2003), \href{http://arxiv.org/abs/hep-ph/0211149}{{\ttfamily
  hep-ph/0211149}}.

\bibitem{Kanazawa:2014fla}
T.~Kanazawa and A.~Yamamoto, {\em Phys. Rev.} {\bf D91},   074508  (2015),
  \href{http://arxiv.org/abs/1411.4667}{{\ttfamily 1411.4667}}.

\bibitem{Creutz:1979dw}
M.~Creutz, {\em Phys. Rev. Lett.} {\bf 43}, 553  (1979).

\bibitem{Irges:2015uta}
N.~Irges, G.~Koutsoumbas and K.~Ntrekis, {\em Phys. Rev.} {\bf D92},   094506
  (2015), \href{http://arxiv.org/abs/1503.06431}{{\ttfamily 1503.06431}}.

\bibitem{DelDebbio:2013rka}
L.~Del~Debbio, R.~D. Kenway, E.~Lambrou and E.~Rinaldi, {\em Phys. Lett.} {\bf
  B724}, 133  (2013), \href{http://arxiv.org/abs/1305.0752}{{\ttfamily
  1305.0752}}.

\bibitem{Irges:2009qp}
N.~Irges and F.~Knechtli, {\em Phys.Lett.} {\bf B685}, 86  (2010),
  \href{http://arxiv.org/abs/0910.5427}{{\ttfamily 0910.5427}}.

\bibitem{Dimopoulos:2006qz}
P.~Dimopoulos, K.~Farakos and S.~Vrentzos, {\em Phys. Rev.} {\bf D74},   094506
   (2006), \href{http://arxiv.org/abs/hep-lat/0607033}{{\ttfamily
  hep-lat/0607033}}.

\bibitem{Farakos:2008iw}
K.~Farakos and S.~Vrentzos, {\em Phys. Rev.} {\bf D77},   094511  (2008),
  \href{http://arxiv.org/abs/0801.3722}{{\ttfamily 0801.3722}}.

\bibitem{Farakos:2008se}
K.~Farakos and S.~Vrentzos, {\em Phys. Rev.} {\bf D78},   114502  (2008),
  \href{http://arxiv.org/abs/0807.3463}{{\ttfamily 0807.3463}}.

\bibitem{Dvali:1996bg}
G.~R. Dvali and M.~A. Shifman, {\em Nucl. Phys.} {\bf B504}, 127  (1997),
  \href{http://arxiv.org/abs/hep-th/9611213}{{\ttfamily hep-th/9611213}}.

\bibitem{Laine:2004ji}
M.~Laine, H.~Meyer, K.~Rummukainen and M.~Shaposhnikov, {\em JHEP} {\bf 0404},
   027  (2004), \href{http://arxiv.org/abs/hep-ph/0404058}{{\ttfamily
  hep-ph/0404058}}.

\bibitem{Gies:2003ic}
H.~Gies, {\em Phys. Rev.} {\bf D68},   085015  (2003),
  \href{http://arxiv.org/abs/hep-th/0305208}{{\ttfamily hep-th/0305208}}.

\bibitem{Morris:2004mg}
T.~R. Morris, {\em JHEP} {\bf 01},   002  (2005),
  \href{http://arxiv.org/abs/hep-ph/0410142}{{\ttfamily hep-ph/0410142}}.

\bibitem{Beard:1997ic}
B.~B. Beard {\em et~al.}, {\em Nucl. Phys. Proc. Suppl.} {\bf 63}, 775  (1998),
  \href{http://arxiv.org/abs/hep-lat/9709120}{{\ttfamily hep-lat/9709120}}.

\bibitem{Itou:2014iya}
E.~Itou, K.~Kashiwa and N.~Nakamoto  (2014),
  \href{http://arxiv.org/abs/1403.6277}{{\ttfamily 1403.6277}}.

\bibitem{Rinaldi:aa}
E.~Rinaldi, {Non--perturbative aspects of physics beyond the Standard Model},
  PhD thesis, Edinburgh U.  (2013).

\bibitem{Chandrasekharan:1996ih}
S.~Chandrasekharan and U.~J. Wiese, {\em Nucl. Phys.} {\bf B492}, 455  (1997),
  \href{http://arxiv.org/abs/hep-lat/9609042}{{\ttfamily hep-lat/9609042}}.

\bibitem{Brower:1997ha}
R.~Brower, S.~Chandrasekharan and U.~Wiese, {\em Phys.Rev.} {\bf D60},   094502
   (1999), \href{http://arxiv.org/abs/hep-th/9704106}{{\ttfamily
  hep-th/9704106}}.

\bibitem{Irges:2012tu}
N.~Irges and G.~Koutsoumbas, {\em JHEP} {\bf 08},   103  (2012),
  \href{http://arxiv.org/abs/1205.0379}{{\ttfamily 1205.0379}}.

\bibitem{Hosotani:1983vn}
Y.~Hosotani, {\em Phys.Lett.} {\bf B129},   193  (1983).

\bibitem{Cossu:2013ora}
G.~Cossu, H.~Hatanaka, Y.~Hosotani and J.-I. Noaki, {\em Phys.Rev.} {\bf D89},
   094509  (2014), \href{http://arxiv.org/abs/1309.4198}{{\ttfamily
  1309.4198}}.

\bibitem{Cossu:2009sq}
G.~Cossu and M.~D'Elia, {\em JHEP} {\bf 07},   048  (2009),
  \href{http://arxiv.org/abs/0904.1353}{{\ttfamily 0904.1353}}.

\bibitem{Kovtun:2007py}
P.~Kovtun, M.~Unsal and L.~G. Yaffe, {\em JHEP} {\bf 06},   019  (2007),
  \href{http://arxiv.org/abs/hep-th/0702021}{{\ttfamily hep-th/0702021}}.

\bibitem{deForcrand:2002hgr}
P.~de~Forcrand and O.~Philipsen, {\em Nucl. Phys.} {\bf B642}, 290  (2002),
  \href{http://arxiv.org/abs/hep-lat/0205016}{{\ttfamily hep-lat/0205016}}.

\bibitem{Akerlund:2015poa}
O.~Akerlund and P.~de~Forcrand, {\em PoS} {\bf LATTICE2014},   272  (2015),
  \href{http://arxiv.org/abs/1503.00429}{{\ttfamily 1503.00429}}.

\bibitem{Kubo:2001zc}
M.~Kubo, C.~Lim and H.~Yamashita, {\em Mod.Phys.Lett.} {\bf A17}, 2249  (2002),
  \href{http://arxiv.org/abs/hep-ph/0111327}{{\ttfamily hep-ph/0111327}}.

\bibitem{Irges:2006zf}
N.~Irges and F.~Knechtli  (2006),
  \href{http://arxiv.org/abs/hep-lat/0604006}{{\ttfamily hep-lat/0604006}}.

\bibitem{Irges:2012mp}
N.~Irges, F.~Knechtli and K.~Yoneyama, {\em Phys.Lett.} {\bf B722}, 378
  (2013), \href{http://arxiv.org/abs/1212.5514}{{\ttfamily 1212.5514}}.

\bibitem{Antoniadis:2001cv}
I.~Antoniadis, K.~Benakli and M.~Quiros, {\em New J.Phys.} {\bf 3},  ~20
  (2001), \href{http://arxiv.org/abs/hep-th/0108005}{{\ttfamily
  hep-th/0108005}}.

\bibitem{Alberti:2015pha}
M.~Alberti, N.~Irges, F.~Knechtli and G.~Moir, {\em JHEP} {\bf 09},   159
  (2015), \href{http://arxiv.org/abs/1506.06035}{{\ttfamily 1506.06035}}.

\bibitem{Arnold:2002jk}
G.~Arnold, B.~Bunk, T.~Lippert and K.~Schilling, {\em Nucl.Phys.Proc.Suppl.}
  {\bf 119}, 864  (2003),
  \href{http://arxiv.org/abs/hep-lat/0210010}{{\ttfamily hep-lat/0210010}}.

\end{thebibliography}

\end{document}